\documentstyle[12pt,epsf]{article}

\newcommand{\reseteqnum}{\setcounter{equation}{0}}

\newcommand{\ba}{\begin{array}}
\newcommand{\ea}{\end{array}}
\newcommand{\nn}{\nonumber \\}
\newcommand{\be}{\begin{equation}}
\newcommand{\ee}{\end{equation}}
\newcommand{\bea}{\begin{eqnarray}}
\newcommand{\eea}{\end{eqnarray}}
\newcommand{\beann}{\begin{eqnarray*}}
\newcommand{\eeann}{\end{eqnarray*}}
\newcommand{\bd}{\begin{description}}
\newcommand{\ed}{\end{description}}

\newcommand{\bra}[1]{\langle #1 \vert}
\newcommand{\ket}[1]{\vert #1 \rangle}

\newcommand{\bbra}[1]{\langle\kern-3.5pt\langle #1 \vert}
\newcommand{\kket}[1]{\vert #1 \rangle\kern-3.5pt\rangle}

\renewcommand{\slash}[2]{#1\kern-#2pt\mbox{\it/}}
\newcommand{\sle}{\slash{\epsilon}{5.2}}
\newcommand{\slp}{\slash{p}{6.0}}
\newcommand{\slq}{\slash{q}{5.8}}

\newcommand{\slqh}{\slash{\widehat{q}}{5.8}}
\newcommand{\Sp}[1]{\frac{1}{4}{\rm tr}\left[\: #1 \:\right]}

\newcommand{\gmu}{\gamma^\mu}

\newcommand{\gnu}{\gamma^\nu}

\newcommand{\tr}{{\rm tr}}

\newcommand{\qh}{{\widehat{q}}}
\newcommand{\bq}{{\mbox{\boldmath $q$}}}

\newcommand{\lqcd}{\Lambda_{\rm QCD}}

\newcommand{\SI}{S_F^{-1}}

\newcommand{\intdk}{\int\!\!\frac{d^4k}{(2\pi)^4i}}
\newcommand{\intdp}{\int\!\!\frac{d^4p}{(2\pi)^4i}}

\newcommand{\T}{{\rm T}}
\newcommand{\psib}{\overline\psi}
\newcommand{\bV}{\langle{V(q,\epsilon)}\vert}
\newcommand{\kV}{\vert{V(q,\epsilon)}\rangle}
\newcommand{\argp}{(p;q,\epsilon)}
\newcommand{\argk}{(k;q,\epsilon)}
\newcommand{\mvn}{M_V^{(n)}}

\newcommand{\lan}{\lambda_{n'}}
\newcommand{\lon}{\lambda_{1'}}
\newcommand{\ltw}{\lambda_{2'}}
\newcommand{\lth}{\lambda_{3'}}
\newcommand{\chib}{\overline\chi}

\catcode`\@=11
\newbox\tempboxa
\newdimen\captionboxsubcount
\def\capsize#1{\captionboxsubcount=#1pt}
\newdimen\captionboxsub
\captionboxsub=\hsize \advance\captionboxsub by -\captionboxsubcount
\advance\captionboxsub by -\captionboxsubcount
\long\def\@makecaption#1#2{
 \setbox\@tempboxa\hbox{#1: #2}
 \ifdim \wd\@tempboxa >\captionboxsub
\rightskip=\captionboxsubcount \leftskip=\captionboxsubcount #1: #2
\else \hbox to\hsize{\hfil\box\@tempboxa\hfil}
 \fi}
\catcode`\@=12
\capsize{30}

\begin{document}

\begin{titlepage}

\begin{flushright}
\ \hfill
\begin{tabular}{l}
KUNS-1304 \\
HE(TH)~94/16 \\
SU-4240-592 \\
hep-ph/9505206 \\
\end{tabular}
\end{flushright}

\bigskip
\begin{center} \LARGE \bf
Solving the Homogeneous \\
Bethe-Salpeter Equation \\
\end{center}
\bigskip

\begin{center} \Large
Masayasu Harada${}^{a,}$\footnote{
e-mail address :
{\tt mharada@npac.syr.edu}}
and
Yuhsuke Yoshida${}^{b,}$\footnote{e-mail address :
{\tt yoshida@gauge.scphys.kyoto-u.ac.jp}} \\
\end{center}

\begin{center} \large \it
${}^a$~
Department of Physics, Syracuse University \\
Syracuse, NY 13244-1130, USA \\

${}^b$~
Department of Physics, Kyoto University \\
Kyoto 606-01, Japan \\
\end{center}

\begin{center}
May 1, 1995
\end{center}

\smallskip

\begin{abstract}
\normalsize
We study a method for solving the homogeneous Bethe-Salpeter
equation.
By introducing a `fictitious' eigenvalue $\lambda$ the homogeneous
Bethe-Salpeter equation is interpreted as a linear eigenvalue
equation, where the bound state mass is treated as an input
parameter.
Using the improved ladder approximation with the constant fermion
mass, we extensively study the spectrum of the fictitious eigenvalue
$\lambda$ for the vector bound states and find the discrete spectrum
for vanishing bound state mass.
We also evaluate the bound state masses by tuning appropriate
eigenvalues $\lambda$ to be unity, and find massless vector bound
states for specific values of the constant fermion
masses.

\end{abstract}

\end{titlepage}


\section{Introduction
\label{Introduction}}
\reseteqnum

Bethe-Salpeter (BS) equations\cite{Nambu,BS,Nakanishi}
for the fermion-antifermion systems in gauge theories are important
for studying the properties of positronium, mesons and so on.
The BS equation in the improved ladder approximation was applied to
the bound state problem in QCD.
Especially, the numerical solution for the massless pion was given in
refs.~\cite{ABKMN,JM}, and they succeeded to reproduce the property of
the chiral symmetry in the light quark sector.
Moreover, the BS equation for the $B$ meson was solved in the heavy
quark limit.\cite{KMY}

Let us explain the solvability of the homogeneous BS equation
for the pion case.
The reason is essentially found in the Nambu-Goldstone's
theorem\cite{N,Goldstone}.
First of all, the pion is known to be massless without solving the BS
equation.
Moreover, thanks to the axial Ward-Takahashi identity,
one scalar component, $\widehat S$, of
the pion BS amplitude is found to be
identical with the mass function of the quark.
Then, the other components are obtained by solving the resultant
inhomogeneous equation,
where the component $\widehat S$ is treated as
an inhomogeneous term and linear algebraic
techniques are applicable.

The reason why the homogeneous BS equation for the $B$ meson can be
solved is that the equation in the heavy quark limit becomes an
eigenvalue equation.
The BS amplitude and the binding energy of the $B$ meson are
corresponding to the eigenvector and the eigenvalue, respectively.
Thus, the problem is reduced to solving the eigenvalue equation, which
can be done by linear algebraic techniques.

How about other cases?
Generally, there are two problems for solving the homogeneous BS
equation for the massive bound state:
\begin{itemize}
\item[i)]
We need the fermion propagator in solving the BS equation.
It is natural to take account of the quantum corrections to the
fermion propagator.
Actually this is required by the consistency with the chiral
Ward-Takahashi
identity\cite{Maskawa-Nakajima,Kugo-Mitchard,BandoHaradaKugo} in QCD.
We have to work in the time-like region for the center-of-mass
momentum $q^\mu$ (on-shell),
while we carry out the Wick rotation to the relative momentum $p^\mu$
(off-shell).
In a certain gauge choice, the fermion propagator in the homogeneous
BS equation takes the form
$iS_F^{-1}(p\!\pm q/2) = \slp\pm \slq/2 - \Sigma(-(p\!\pm q/2)^2)$,
where $q^2$ is identical to the
bound state mass squared and the time
component $p^0$ is pure imaginary.
The momenta flowing along the fermion and antifermion
lines become complex.
Thus, we need the mass function $\Sigma(z)$ on the complex plane.
\item[ii)]
The homogeneous BS equation depends on the bound state mass
complicatedly.
It is not any linear eigenvalue equation so that it seems
impossible to obtain the bound state mass and its corresponding BS
amplitudes directly.
Even when we adopt the constant mass approximation for the mass
function, the complexity is not avoided.
The homogeneous BS equation includes both the linear and
quadratic terms of the bound state mass simultaneously.
\end{itemize}

One way to {\em avoid} the above two problems
is using the {\em inhomogeneous} BS equation in space-like
region.\cite{AKM,HaradaYoshida}
In general, the solution of the inhomogeneous BS equation has the
perturbative expansion series which begins with the inhomogeneous
part.
The inhomogeneous BS equation may be solved by the iterations:
the inhomogeneous part is taken as a first trial solution, and is
substituted into the inhomogeneous BS equation in order to make more
accurate solution.
Repeating this iteration many times, we obtain the solution with
appropriate accuracy.

Another way is taking further approximation in addition to the ladder
approximation.
When we take the BS kernel to be the instantaneous (Coulomb-like)
interaction, the homogeneous BS equation reduces to the Salpeter
equation.\cite{Salpeter}\
On the other hand, in ref.~\cite{MJ} the authors
work for the Euclidean total momentum and solve the BS equation
by expanding the BS
amplitude in terms of $SO(4)$ orthogonal Tschebyshev polynomials and
dropping higher order polynomials.
They investigate various mesons with their approximation.\cite{JM2}

Now, let us ask the following question:
``How do we {\em solve} the above two problems i) and ii)
to obtain solutions of the {\em homogeneous} BS equation?''
As for the problem i),
the Schwinger-Dyson (SD) equation on the complex plane was solved to
obtain the mass function.\cite{FukudaKugo,AtkinsonBlatt} This analytic
continuation is also done using the SD equation in the integral form.
[In appendix~\ref{SD}, following ref.~\cite{KY},
we briefly review how to derive the fermion mass
function on the complex plane.]
Then we concentrate our attention on the problem ii).

Traditionally,
the BS equation is solved by regarding a coupling constant as an
eigenvalue and the bound state mass as an input.\cite{Nakanishi}
In this paper, to solve the homogeneous BS equation for a given
coupling constant numerically, we introduce a `fictitious' eigenvalue
$\lambda$ by replacing the BS kernel $K$ with $\lambda K$.
We notice that it is corresponding to replacing the coupling
$\alpha$ with $\lambda\alpha$ in the (improved) ladder and constant
fermion mass approximations.
By adjusting the input
bound state mass so as to set the fictitious
eigenvalue $\lambda$ equal to unity, we get masses of the bound
states.
As stated above, it is natural to use the full propagator which is
determined by the SD equation in the same approximation as that
for the BS equation.
But here we concentrate our attention on the numerical method to solve
the homogeneous BS equation, and we adopt the constant fermion mass
approximation.

In many literatures\cite{Seto},
by regarding the coupling constant as an eigenvalue,
the BS equation is solved for the fermion-antifermion bound state in
the fixed coupling case.
The behavior of the `eigenvalue' $\alpha$, which corresponds to our
fictitious eigenvalue $\lambda$, is studied in detail.
For applying to QCD we should use the running
coupling\cite{Higashijima,Miransky},
but
we do not know the behavior of the fictitious eigenvalue
$\lambda$.
Then, before calculating the bound state mass,
we study the behavior of $\lambda$ relating it with the
norms of BS amplitudes.
Moreover, we check the existence of the discrete spectrum:
our choice of the running coupling\cite{HaradaYoshida}
becomes a constant in the low energy region and the low energy
behavior of the theory would be similar to that of the strong coupling
QED.
For a type of BS equation in the strong coupling QED, it is
shown\cite{Seto,Tiktopoulos} that there is no discrete spectrum.

After studying the behavior of $\lambda$,
we solve the original BS equation for the  vector bound state
numerically, and obtain the bound state mass.
As an instructive example to check the validity of our method,
we also calculate the spectrum of the orthopositronium,
which is a vector bound state of electron and positron.

This paper is organized as follows.
In section~\ref{HBSE} we review the homogeneous BS equation.
The BS amplitude is defined and is expanded in terms of the
invariant amplitudes.
The normalization condition for the BS amplitude is given.
Section~\ref{MSHBSE} is devoted to explanation of the method for
solving the homogeneous BS equation.
We extensively study the behavior of the fictitious eigenvalue
$\lambda$ in the running coupling case
in section~\ref{sec: spec}.
In section~\ref{sec: mass}, as a consistency check we study the
positronium in the weak coupling QED, and we calculate the
bound state masses in the running coupling case.
The discussions are found in section~\ref{Summary and Discussion}.

\section{Homogeneous Bethe-Salpeter Equation}
\label{HBSE}
\reseteqnum

In this section we show the basic formulations in solving the
homogeneous Bethe-Salpeter (HBS) equation for vector bound state.
Our formulations are rather transparent especially for the vector
case.

\subsection{BS Amplitude}

Let us express the fermion-antifermion bound state of the vector type
($J^{PC}=1^{--}$) as $\kV$ where $q_\mu$ is the
momentum of the bound state and
$\epsilon_\mu$ is the polarization vector. [$\epsilon\!\cdot\! q = 0$,
$\epsilon^2=-1$.]
The BS amplitude $\chi(p;q,\epsilon)$ is defined by
\be
\bra{0} \T \psi(x)\psib(y)\kV = e^{-iqX}
\int\!\!\frac{d^4p}{(2\pi)^4}~ e^{-ipr}~ \chi(p;q,\epsilon) ~,
\label{eq: def chi}
\ee
where $X^\mu$ and $r^\mu$ are the center-of-mass coordinate and the
relative coordinate, respectively:
\be
X^\mu = \frac{x^\mu + y^\mu}{2}~, \qquad r^\mu = x^\mu - y^\mu ~.
\ee
The BS amplitude is bispinor and is defined only on the mass-shell of
the bound state $q^2 = M_V^2$, where $M_V$ is the mass of the bound
state.
We usually perform the Wick rotation to the BS amplitude, and the time
component of the relative momentum $p_\mu$ becomes pure imaginary,
while the quantities $q_\mu$ and $\epsilon_\mu$ are Minkowski vectors
and real.

The BS amplitude of the vector bound state is expanded by eight
invariant amplitudes $\chi^1$, $\cdots$, $\chi^8$:
\be
\chi\argp = \sum_{i=1}^8 \Gamma_i\argp\, \chi^i(p;q) ~.
\ee
We note that the dependence of the polarization vector $\epsilon_\mu$
is isolated in the bispinor base $\Gamma_i\argp$.
The following choice of the bispinor bases $\Gamma_i\argp$
($i=1,\cdots,8$) are convenient:
\be
\ba{llll}
\Gamma_1 = \epsilon\kern-5.2pt\mbox{\it/} , &
\Gamma_2 = \frac{1}{2}[\sle,\slp](p\cdot\qh) , &
\Gamma_3 = \frac{1}{2}[\sle,\slqh] , &
\Gamma_4 = \frac{1}{3!}[\sle, \slp, \slqh] ,\bigskip\\
\Gamma_5 = (\epsilon\cdot p) , &
\Gamma_6 = \slp(\epsilon\cdot p) , &
\Gamma_7 = \slqh (p\cdot\qh)(\epsilon\cdot p) , &
\Gamma_8 = \frac{1}{2}[\slp, \slqh](\epsilon\cdot p) , \bigskip
\end{array} \label{eq: bispinor base}
\ee
where $\widehat{q}_\mu = q_\mu/M_V$ and $[a,b,c] \equiv
a[b,c] + b[c,a] + c[a,b]$.
The invariant amplitude $\chi^i(p;q)$ is a scalar function in $p^2$
and $p\cdot q$.%
\footnote{
We are considering the bound states which have discrete spectrum.
The quantity $q^2 = M_V^2$ labels the degree of the radial excitation
if we introduce it as $\chi=\chi(p^2,p\cdot q,q^2)$.
}
So, after the Wick rotation, it is convenient to introduce the real
variables $u$ and $x$ by
\be
p\cdot q = iM_V\,u, \qquad p^2 = -u^2 -x^2 ~.
\ee
The time component of $p^\mu$ is $iu$ and the magnitude of the spatial
components of $p^\mu$ is $x$ in the rest frame
$q^\mu=(M_V,\vec0)$.
The invariant amplitude is a function in $u$ and $x$; $\chi^i =
\chi^i(u,x)$.
By the charge conjugation properties
\bea
C\chi(-p;q,\epsilon)^TC^{-1} &=& - \chi\argp ~,\nn
C\Gamma_i(-p;q,\epsilon)^TC^{-1} &=& - \Gamma_i\argp ~,
\label{eq: cc}
\eea
where $C = i\gamma^2\gamma^0$, all the invariant amplitudes
$\chi^i(u,x)$ are found to be even functions in
$u$:\cite{HaradaYoshida}
\be
\chi^i(u,x) = \chi^i(-u,x) ~. \label{eq: even function}
\ee

\subsection{Homogeneous BS Equation}
The HBS equation reads
\be
T\,\chi = K\,\chi ~. \label{eq: HBS1}
\ee
\begin{figure}[bhtp]
\begin{center}
\ \epsfbox{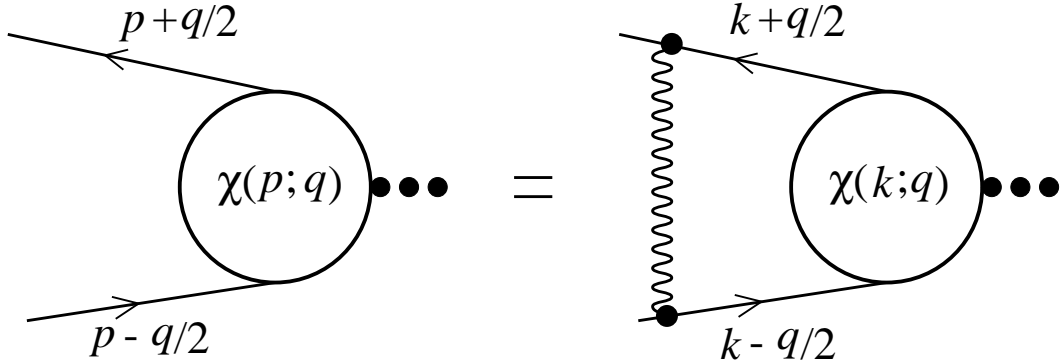}
\vspace{-5pt}
\caption[]{
The Feynman diagram of the homogeneous BS equation (\ref{eq: HBS1}) in
the (improved) ladder approximation.
The quark lines in the RHS of the equation are truncated.
}
\label{fig: HBS}
\end{center}
\end{figure}
The diagram for the HBS equation in the (improved) ladder
approximation is shown in fig.~\ref{fig: HBS}.
The kinetic part $T$ is given by
\be
T(p;q) = \SI(p+q/2)\otimes\SI(p-q/2) \nn
\label{eq: T}
\ee
with the tensor product notation
\be
(A \otimes B) \chi = A \chi B ~.
\ee
In this approximation, the BS kernel $K$ is given by
\be
K(p,k) = C_2 \frac{g^2(p,k)}{-(p-k)^2}
\left(
 g_{\mu\nu}
 - \frac{(p-k)_\mu (p-k)_\nu}{(p-k)^2}
\right) \gmu \otimes \gnu ~.
\label{eq: K}
\ee
In eq.~(\ref{eq: HBS1}) we use the inner product rule
\be
K\chi(p;q) = \intdk K(p,k)\chi(k;q) ~.
\ee
In the improved ladder approximation we use the running coupling,
the asymptotic form of which is given by
\be
\alpha(\mu^2) \equiv \frac{g^2(\mu^2)}{4\pi}
{}~~ \mathop{\longrightarrow}_{\mu\rightarrow\infty} ~~
\frac{\alpha_0}{\ln\mu^2} \ ,
\label{eq: asymptic form}
\ee
where $\alpha_0 = 12\pi/(11N_c - 2N_f)$ with $N_c$ and
$N_f$ being the number of colors and flavors, respectively.

The HBS equation (\ref{eq: HBS1}) determines the eight invariant
amplitudes $\chi^1$, $\cdots$, $\chi^8$ up to the overall
normalization constant as well as the
bound state mass $M_V$.
Multiplying eq.~(\ref{eq: HBS1}) by the conjugate bispinor base
$\overline\Gamma_i\argp \equiv
\gamma_0\Gamma_i(p^*;q,\epsilon)^\dagger \gamma_0$,
taking the trace of spinor indices and summing over the polarizations,
we finally obtain
\be
T_{ij}(u,x)\, \chi^j(u,x) =
\int\!\frac{y^2dydv}{8\pi^3}~K_{ij}(u,x;v,y)\, \chi^j(v,y) ~,
\label{eq: HBS2}
\ee
where the summation over the repeated indices is promised and
\bea
T_{ij}(u,x) &=& \sum_\epsilon ~ \Sp{ \overline\Gamma_i\argp
T(p;q) \Gamma_j\argp } ~,\nn
K_{ij}(u,x;v,y) &=& \int_{-1}^{1}\!\!d\cos\theta \sum_\epsilon ~
\Sp{ \overline\Gamma_i\argp K(p,k) \Gamma_j\argk } ~.
\label{def:mat:T K}
\eea
We omit the explicit forms of $T_{ij}(u,x)$ and $K_{ij}(u,x;v,y)$
here because they are somewhat complicated.
Due to the charge conjugation property (\ref{eq: cc}) of the bispinor
base, the $8\times8$ matrices $T_{ij}$ and $K_{ij}$ are real and
self-conjugate:
\be
\ba{lllll}
T_{ij}(u,x)^\dagger &=& T_{ij}(u,x)^T &=& T_{ji}(u,x) ~, \\
K_{ij}(u,x;v,y)^\dagger &=& K_{ij}(u,x;v,y)^T &=& K_{ji}(v,y;u,x) ~.
\label{eq: TK matrices}
\ea
\ee
The complex conjugate and the transpose are taken in the $8\times 8$
space.
Using the property (\ref{eq: even function}), we are allowed to
restrict the integral region of $v$ to
positive, $v>0$.
The BS kernel in eq.~(\ref{eq: HBS2}) is replaced such that
\be
\int\!\!dv\, K_{ij}(u,x;v,y)\chi^j(v,y)
= \mathop{\int}_{v>0}\!\!dv
\left[\,K_{ij}(u,x;v,y) + K_{ij}(u,x;-v,y)
\,\right]\chi^j(v,y) ~.
\label{eq: prescription1}
\ee
Thus, we can treat all the variables $u$, $x$, $v$, $y$ as positive
values.

In order to solve eq.~(\ref{eq: HBS2}) numerically we discretize
the momentum spaces $(u,x)$ and $(v,y)$.
The effective parameterizations for the variables $u$ and $x$ are
\be
u = e^U ~, \qquad x = e^X ~,\label{eq: UX}
\ee
and similarly for $v$ and $y$.
These new variables are discretized at $N_{BS}$ points evenly spaced
in the intervals
\bea
U , V \in \left[\lambda_U,\Lambda_U\right] ~, \nonumber\\
X , Y \in \left[\lambda_X,\Lambda_X\right] ~.
\label{cutoffs} \label{eq: cutoffs}
\eea
Then, the momentum integration becomes the summation
\be
\mathop{\int}_{v>0}\!\!y^2dydv \cdots \qquad\longrightarrow\qquad
D_UD_X \sum_{V,Y} VY^3 \cdots ~,
\ee
where
\bea
D_U = (\Lambda_U-\lambda_U)/(N_{BS}-1) ~, \nonumber\\
D_X = (\Lambda_X-\lambda_X)/(N_{BS}-1) ~.
\eea

The BS kernel $K_{ij}(u,x;v,y)$ has an integrable singularity at
$(u,x) = (v,y)$ which comes from the pole in the propagator of the
gauge boson (cf. eq.~(\ref{eq: K})).
In order to avoid this singularity we regularize it by taking the
four-point splitting\cite{AKM}
\bea
K_{ij}(u,x;v,y) &\rightarrow& \frac{1}{4}\Big[\,
K_{ij}(u,x;v_+,y_+) + K_{ij}(u,x;v_+,y_-)  \nn
&& + K_{ij}(u,x;v_-,y_+) + K_{ij}(u,x;v_-,y_-)
\,\Big] ~,\label{eq: prescription2}
\eea
where $v_\pm = \exp(V \pm D_U/4)$ and $y_\pm = \exp(Y \pm D_X/4)$.

Now, we are ready to solve eq.~(\ref{eq: HBS2}) numerically.
The HBS equation (\ref{eq: HBS2}) becomes a finite dimensional linear
equation in terms of the invariant BS amplitude $\chi^i$.
All we have to do is to obtain simultaneously the bound state
mass $M_V$ and the corresponding BS amplitude $\chi$.

\subsection{Normalization Condition}

First, we introduce the
conjugate BS amplitude $\overline\chi(p;q,\epsilon)$ as
\be
\bV\T \psi(y)\psib(x)\ket{0} = - e^{iqX}
\int\!\!\frac{d^4p}{(2\pi)^4}~ e^{ipr}~ \overline\chi(p;q,\epsilon) ~.
\ee
The relation between the conjugate BS amplitude
and the BS amplitude is given by
\be
\overline\chi\argp =
\gamma_0\left[\chi(p^*;q,\epsilon)\right]^\dagger\gamma_0 ~.
\ee
Now, let us fix the normalization of the BS amplitude $\chi$.
Usually the bound state $\kV$ is normalized by
\be
\langle V(q',\epsilon')|V(q,\epsilon)\rangle = 2q_0(2\pi)^3\,
\delta_{\epsilon,\epsilon'}\delta^3(\bq-\bq') ~.\label{eq: inv nor}
\ee
This condition fixes the normalization of $\chi$ via its
definition (\ref{eq: def chi}).
The convenient form for the normalization of $\chi$ is given by the
Mandelstam formula
\be
  \bbra{\chi}
    \frac{\partial T}{\partial q^\mu}
  \kket{\chi}
  = 2q_\mu ~.
\label{eq: Mandelstam}
\ee
Here we use the bra-ket notation\cite{KMY}
\be
\bbra{\chi}A\kket{\psi} =
\intdp \tr \left[ \chib(p;q) (A\psi)(p;q)\right]
\ee
with the trace taken over the spinor indices and all the other
inner degrees of freedom.
Multiplying eq.~(\ref{eq: Mandelstam}) by $q^\mu$ and using
$q^\mu(\partial/\partial q^\mu) = M_V(\partial/\partial M_V)$ we get
\be
2M_V =
  \bbra{\chi}
    \frac{\partial T}{\partial M_V}
  \kket{\chi} ~.
\label{eq: amplitude norm}
\ee
It is easy to find the normalization condition in component form as
\be
2M_V = \frac{N_c}{2\pi^3}\int\!\!x^2dxdu~
(\chi^i(-u,x))^*\frac{\partial T_{ij}}{\partial M_V} \chi^j(u,x) ~.
\label{eq: amp norm}
\ee

\section{The Method for Solving the Homogeneous ~~~\mbox{}
Bethe-Salpeter Equation}
\label{MSHBSE}
\reseteqnum

In this section we study the method for solving the homogeneous
Bethe-Salpeter (HBS) equation and explain how to apply it to the
numerical calculation.

Let us start with writing down the HBS equation (\ref{eq: HBS1}):
\be
T\,\chi = K\,\chi ~. \label{eq: HBS1.b}
\ee
This HBS equation simultaneously determines the mass spectrum
$M_V^{(1)}$, $M_V^{(2)}$, $M_V^{(3)}$, $\cdots$ of
the bound states and the BS amplitudes $\chi_1\argp$, $\chi_2\argp$,
$\chi_3\argp$, $\cdots$ up to their normalizations.
Therefore, to solve the HBS equation is to find a mass $M_V$
of a certain bound state and its corresponding BS amplitude
$\chi\argp$.
However, it is difficult to solve HBS equation (\ref{eq: HBS1.b}) as
it is.
For definiteness, we choose the suffix $n$ as $M_V^{(1)} \le$
$M_V^{(2)} \le$ $M_V^{(3)} \le \cdots$.

Usually, the coupling constant $\alpha$ is regarded as an eigenvalue
of HBS equation (\ref{eq: HBS1.b}) for a given bound state
mass,\cite{Nakanishi}
where we use the ladder and the constant mass approximations
simultaneously.
The dependence on the coupling constant is isolated in the BS kernel
and the coupling factors out.
As far as we adopt those approximations,
this is extended to the running coupling case by regarding
the overall constant $\alpha_0$ given in eq.~(\ref{eq: asymptic form})
as an eigenvalue.
On the contrary, if we consistently use the full propagator determined
by the SD equation in the same approximation as that for the BS
equation, the kinetic part $T$ also depends on the coupling $\alpha$
as well as the BS kernel $K$ does.
Then, the HBS equation (\ref{eq: HBS1.b}) cannot be regarded
as a simple linear eigenvalue equation for the coupling.

The original idea to regard the HBS equation as a linear eigenvalue
equation for the coupling and the BS amplitude is easily extended to
generic cases.
By introducing a `fictitious' eigenvalue $\lambda$ we interpret the
HBS equation (\ref{eq: HBS1.b}) as a linear eigenvalue equation for a
given bound state mass $M_V$:
\be
\left(\frac{1}{\lambda}\right)
T \chi = K \chi ~, \label{eq: eigen HBS}
\ee
where the kinetic part $T$ is regarded as a weight function.
We calculate the bound state mass for a given coupling
$\alpha$ numerically from this equation.
When we consider the mass $M_V$ as an input parameter, the form of
eq.~(\ref{eq: eigen HBS}) guarantees the solvability: we can solve
eq.~(\ref{eq: eigen HBS}) by standard linear algebraic techniques for
eigenvalue equations, and obtain the solutions of the eigenvalue
$\lambda$.
The fictitious eigenvalue is essentially the same as the coupling
constant, $\lambda \propto \alpha$, when we use both the (improved)
ladder and constant fermion mass approximations.
The resultant HBS equation (\ref{eq: eigen HBS}) simultaneously
describes various systems parameterized by the coupling constant
$\alpha$.

In what follows we first study the basic notions of the eigenvalue
equation (\ref{eq: eigen HBS}) and then we explain the method for
solving the original HBS equation (\ref{eq: HBS1.b}).
We note that we have to check the existence of the discrete spectrum
of $\lambda$.
As we will show in the next section, there exists the discrete
spectrum in the ladder and constant mass approximations.
Here and henceforth we consider such case.
In order to distinguish the quantum numbers of the eigenvalue
equation (\ref{eq: eigen HBS}) from that of the original HBS equation
(\ref{eq: HBS1.b}) we attach primes to them like $n'$ or $1'$,
$2'$, $3'$, $\cdots$ when we solve eq.~(\ref{eq: eigen HBS}) to obtain
the spectrum of $\lambda$ for a given $M_V$.
The obtained eigenvalue $\lambda$ is a function of the
bound state mass parameter $M_V$, so we write the dependence
explicitly as $\lan(M_V)$.
For the convenience of the numerical calculation we define the
numbering of $\{\lan(M_V)\}$ so as to satisfy
\be
0 \le \lon(M_V) \le \ltw(M_V) \le \lth(M_V) \le \cdots
{}~,\label{eq: ordering}
\ee
where the equality holds when the eigenvalues are degenerate.
In our numerical calculation we encounter the accidental degeneracy of
the two eigenvalues at some point $M_V$.
We call this phenomena `level crossing'.

There are two kinds of norms of the BS amplitude in
eq.~(\ref{eq: eigen HBS}).
The quantity $\bbra{\chi_{n'}}\partial T/\partial M_V
\kket{\chi_{n'}}$, which we call $T'$-norm, gives the normalization
for the BS amplitude as in eq.~(\ref{eq: amplitude norm}).
The non-positive $T'$-norm states are ghosts and cannot be normalized.
So, we consider only the states which have positive $T'$-norm.
The quantity $\bbra{\chi_{n'}}(-T)\kket{\chi_{n'}}$, which we call
$T$-norm, gives also a natural norm for the eigenvalue equation
(\ref{eq: eigen HBS}) in the following sense.
In the case where the eigenvalue $\lambda$ is not always real,
it is easy to derive the well-known relation in linear algebra:
\be
(1/\lambda_{n'} - 1/\lambda_{m'}^*)
\bbra{\chi_{m'}} (-T)\kket{\chi_{n'}} = 0 ~.\label{eq: orthogonality}
\ee
This relation says that  the eigenvalue $\lambda$ is real otherwise
the corresponding state $\chi$ has zero $T$-norm, and
the state $\kket{\chi_{n'}}$ can form a complete orthonormal
set under the inner product $(\chi,\psi) \equiv
\bbra{\chi}(-T)\kket{\psi}$.

For the purpose that we find the solutions with $\lambda=1$, it is
enough to consider $\lambda$ is real and positive.
[We do not consider the possibility that $\lambda$ becomes negative,
because it implies the repulsive interaction which cannot form any
bound states.]
Then, we throw away all non-real eigenvalues and its corresponding
eigenvectors.
Further, we do not consider the complex $M_V$ in solving
eq.~(\ref{eq: eigen HBS}).
If $M_V$ is complex, the corresponding state is complex ghost and is
physically unacceptable.

Now, let us consider the relation between the linear eigenvalue
equation (\ref{eq: eigen HBS}) and the original HBS equation
(\ref{eq: HBS1.b}).
What we obtain from (\ref{eq: eigen HBS}) is the set $\{\,\lan,
\chi_{n'}\,\}$ for a given $M_V$, but what we have to find out is the
set $\{\,\mvn, \chi_n\,\}$.
If we find an eigenvalue
$\lan$ and its corresponding eigenvector
$\chi_{n'}$ such that $\lan = 1$ for a certain value of the mass
$M_V$, this solution also satisfies the original
equation (\ref{eq: HBS1.b}):
\be
0 = (T - \lan K)\chi_{n'} = (T - K)\chi_{n'} ~.
\ee
Namely, the set $\{M_V, \chi_{n'} | \lan=1\}$ is nothing but a
solution of the HBS equation (\ref{eq: HBS1.b}).
Of course, the other states $m' \ne n'$ which do not satisfy
$\lambda_{m'} = 1$ are out of the case.
Then, the physical spectrum $\{\mvn \,|\, n \in {\bf N}\,\}$ of the
bound state masses
is determined by the following set of intrinsic
relations
\be
\lon(M_V) = 1,~ \ltw(M_V) = 1,~ \lth(M_V) = 1,~
\cdots ~.\label{eq: root}
\ee
Clearly, all the roots of these equations give the solutions of the
original HBS equation (\ref{eq: HBS1.b}).

Next, we explain how to input the mass $M_V$ in
eq.~(\ref{eq: eigen HBS}) by a systematic manner for tuning
the eigenvalue to unity ($\lan=1$) numerically.
For this purpose we first consider the differentiability of the
eigenvalue $\lambda(M_V)$.
In general the function $\lan(M_V)$ defined by
eq.~(\ref{eq: ordering}) is not always differentiable
at level crossing point.
We modify the definition of the numbering ($\lan \rightarrow
\lambda_n$) so that the functional forms of the eigenvalues become
always differentiable.
In order to make differentiable form $\lambda(M_V)$ for all
range of $M_V$ we invoke the eigenvalue equation (\ref{eq: eigen HBS})
itself.
Differentiating eq.~(\ref{eq: eigen HBS}) with respect to $M_V$, we
have\cite{Nakanishi}
\be
\frac{d\lambda(M_V)}{d M_V}
= - \lambda(M_V) \frac{
 \bbra{\chi}\frac{\partial T}{\partial M_V} \kket{\chi}}
{\bbra{\chi} (-T) \kket{\chi}} ~, \label{eq: diff}
\ee
where we use the HBS equation for the conjugate BS amplitude.
For normalizable BS amplitudes we put
$\bbra{\chi}\partial T/\partial M_V\kket{\chi} = 2M_V$ according
to eq.~(\ref{eq: amplitude norm}).
The equation (\ref{eq: diff}) implies that if the BS amplitude
$\chi$ is normalizable and have non-vanishing $T$-norm, the
derivative of the corresponding eigenvalue $\lambda$ exists.
Let $\lambda_n(M_V)$ ($n\in {\bf N}$) denote one of such eigenvalues
and let $\chi_n$ denote its corresponding eigenvector.
Once we define the numbering (or we can say tagging) $\lambda_n(M_V)$
at a suitable point $M_V$,
we can uniquely identify the functional form of $\lambda_n(M_V)$
according to the ``differential equation'' (\ref{eq: diff}).
In other words, the differential equation (\ref{eq: diff}) with a
suitable initial condition uniquely determines the form of
$\lambda_n(M_V)$.

Here, we explain the systematic method to tune the eigenvalue.
We first consider the case that the level crossing does not occur, and
we identify $\lambda_n(M_V) = \lan(M_V)$ for $n=1,2,\cdots$.
The relation (\ref{eq: diff}) tells us the response of the eigenvalue
$\lan(M_V)$ for the bound state mass $M_V$.
If we update the bound state mass to $M_V+\delta M_V$, then the
eigenvalue $\lan$ changes to $\lan+\delta M_V(d\lan/dM_V)$ according
to the relation (\ref{eq: diff}).
In order to find the root of the equation $\lan(M_V)=1$ numerically,
we are better to use the Newton's method.
The bound state mass is updated according to
\bea
M_V &\rightarrow& M_V + \delta M_V ~,\nn
\delta M_V &=&
\frac{\bbra{\chi_{n'}}(-T)\kket{\chi_{n'}}}{
\bbra{\chi_{n'}}\frac{\partial T}{\partial M_V}\kket{\chi_{n'}}}
\left(1 - \frac{1}{\lan}\right) ~.
\label{eq: newton}
\eea
When $\lan(M_V)$ is not differentiable at a
level crossing point
$M_0$, $\lan$ and $\chi_{n'}$ in eq.~(\ref{eq: newton})
should be replaced with $\lambda_n$ and $\chi_n$ respectively.
The derivative $d\lan/dM_V$ has two different values at $M_0$
depending on how to take the limits,
 $M_V\searrow M_0$ or $M_V \nearrow M_0$.
If the difference between these two values is not so large,
it is not a problem for the purpose of updating $\lan$:
we do not need the precise value of $M_V$ so as to satisfy
$\lan(M_V)=1$ at one time, whereas we need only more suitable value of
$M_V$ than the previous trial value of $M_V$.
Therefore, we generally use the updating procedure (\ref{eq: newton}).

It is important to study the property of the spectrum of $\lambda$ in
order to know which of the states the solution
$\{ M_V, \chi_{n'} | \lan = 1 \}$ is corresponding to, the ground
state or the the first excited state or the other states.
In this section we assume that the solutions have positive $T$-norm
below.
[Indeed,  our numerical calculations given in the next section
satisfies this assumption.]
Under this assumption, all eigenvalues $\lambda_n(M_V)$
($n=1,2,\cdots$) in consideration are differentiable and decreasing
functions.
Thus, all $\lan(M_V)$'s are also decreasing functions, and each
equation $\lan(M_V)=1$ in eq.~(\ref{eq: root}) has only one root at
most.
Many models suggest that $\lambda(M_V)$ is a decreasing
function (see for example, ref.~\cite{Nakanishi}).
[Smaller value of $M_V$ means larger binding energy which should be
realized by larger coupling to form tight bound state.
$\lambda$ is roughly the same as the coupling.]

We notice that there is a possibility that some of the equations in
(\ref{eq: root}) have no solution, e.g., $\lan(M_V) \ne 1$, for any
real input value $M_V$.
Those $\lambda$'s which are already smaller than unity ($\lambda<1$)
at $M_V=0$ never have the solution of $\lambda(M_V)=1$.
In this case the eigenvalue and its corresponding state
$(\lan,\chi_{n'})$ has no physical correspondence.%
\footnote{
If we allow $M_V$ to take complex value, we may always have roots
of eq.~(\ref{eq: root}).
But, in our calculation we do not consider complex
bound state mass because it is unphysical.
}
In other word, the state $\chi_{n'}$ is annihilated in the physical
states which are determined in terms of eq.~(\ref{eq: root}).
We call this phenomena the ``level annihilation''.
We actually encounter the level annihilation in the running coupling
case, which is studied in the next section.

Now, we will establish the identifications of the
physical states $\{\mvn,\chi_n \}$ with the solutions
$\{ M_V,\chi_{m'}|\lambda_{m'}(M_V)=1\}$.
In the numerical calculation these identifications are easy tasks.
We calculate the set of eigenvalues $\{ \lan \}$ from $M_V=0$ to
a certain large value.
When we use the constant mass and (improved) ladder approximation, the
point $M_V=2m$ is the threshold for the constituent fermions to be
liberated from the bound state.
So, it is enough to calculate $\lambda$ up to $M_V=2m$.
Supposing that there are $l\!-\!1$ level annihilations,
we obtain $\lon\le\cdots\le\lambda_{(l-1)'}<1\le\lambda_{l'}\le\cdots$
at $M_V = 0$.
Then, {\em the $l$-th eigenvalue $\lambda_{l'}$, which becomes unity
first as increasing the input parameter $M_V$ and its corresponding
BS amplitude $\chi_{l'}$ give the ground state.
The $(l\!+\!1)$-th eigenvalue $\lambda_{(l+1)'}$ which becomes unity
second and its corresponding BS amplitude $\chi_{m'}$ give the first
excited state, and so on}.

Finally, we consider the identification of $\lambda_n(M_V)$ with
$\lambda_{m'}(M_V)$ under the level crossing.
When there is no degeneracy between any two eigenvalues of
$\lambda_{m'}$'s, we can uniquely identify the form of
$\lambda_n(M_V)$ according to the continuity: $\lambda_n(M_V) =
\lan(M_V)$ for ${}^\forall n\in{\bf N}$.
[In a model a curve of $\lambda(M_V)$ terminates at some point
$M_V$\cite{Suttorp}, but the continuity of the curve does not spoil.]
However, in general cases there may occur the level crossing.
The level crossing is that the accidental degeneracy of states
for a specific value of the bound state mass $M_V$.
There exist two possible kinds of level crossing:
1) $\lambda_{n'}$ and $\lambda_{m'}$ are degenerate while both are
still differentiable;
2) $\lambda_{n'}$ and $\lambda_{m'}$ are not differentiable
at level crossing point.
Below, we discuss the latter kind of level crossing.
The important thing here is that $\lambda_n(M_V)$ is defined by the
differentiability (\ref{eq: diff}) while $\lan(M_V)$ is defined by the
ordering (\ref{eq: ordering}).
For simplicity here we consider the case when only two states become
degenerate at one time with different slopes of $\lambda(M_V)$.
If one level crossing between the first and the second smallest
$\lambda\,${}'s occurs at $M_V=M_{V1}$ then
$\lambda_n(M_V) = \lan(M_V)$ ($n=1,2,3,\cdots$) for $M_V>M_{V1}$ and
$\lambda_2(M_V)=\lon(M_V)$, $\lambda_1(M_V)=\ltw(M_V)$ and
$\lambda_n(M_V) = \lan(M_V)$ ($n=3,4,\cdots$) for $M_V<M_{V1}$.
This identification is easily understood in the general cases, where
the differentiability of $\lambda_n(M_V)$ uniquely make us identify
its form.
As a result, the phenomena of the level crossing is that two
differentiable curves $\lambda_m(M_V)$ and $\lambda_{n}(M_V)$ cross
at a point $M_V=M_{Vcross}$.

\section{Numerical Calculation}
\reseteqnum

In this section, we first study the functional form of $\lambda(M_V)$
in our numerical method.
We consider the running coupling case (improved ladder approximation)
in the constant fermion mass approximation.
Next, we evaluate the
bound state mass determined by the original HBS
equation eq.~(\ref{eq: HBS1.b}).
This study is motivated by the $\overline q q$ quarkonia in QCD.
Here and henceforth, we rescale all dimensionful quantities by $\lqcd$
otherwise stated.
$\lqcd$ is defined by the blow-up scale of the one-loop running
coupling in the improved ladder model.

Let us consider the important things for the numerical calculations.
As discussed before, in the improved ladder approximation the
asymptotic form of the running coupling is given by the one-loop
renormalization group equation, but we have no idea for the functional
form in the low energy region.
As in ref.~\cite{HaradaYoshida} we use the form of the running
coupling:
\begin{equation}
\alpha(\mu^2) \equiv \frac{g^2(\mu^2)}{4\pi} =
   \alpha_0 \times \left\{\begin{array}{ll}
\displaystyle \frac{1}{t} & \mbox{ if $t_F < t$ } \smallskip\\
\displaystyle \frac{1}{t_F} + \frac{(t_F - t_C)^2
   - (t - t_C)^2}{2t_F^2(t_F - t_C)} &\smallskip
   \mbox{ if $ t_C < t < t_F$ } \\
\displaystyle \frac{1}{t_F} + \frac{(t_F - t_C)}{2t_F^2} &
   \mbox{ if $ t < t_C$ } \smallskip
   \end{array}\right.~,
\label{eq: alpha}
\end{equation}
where $t = \ln \mu^2$ and $\alpha_0 = 12\pi/(11N_c - 2N_f)$ with $N_c$
and $N_f$ being the number of colors and flavors, respectively.
In this paper we fix
$N_c=N_f=3$ and $t_F = 0.5$ and $t_C=2.0$.

In the numerical calculation we have to take care of covering the
supports of both $\bbra{\chi}(-T)\kket{\chi}$ ($T$-norm) and
$\bbra{\chi}\partial T/\partial M_V\kket{\chi}$ ($T'$-norm).
Two supports of the norms should lie within the momentum
cutoffs $[\lambda_U, \Lambda_U]$ and $[\lambda_X, \Lambda_X]$ given in
eq.~(\ref{eq: cutoffs}), otherwise we fail to obtain the precise
values of the eigenvalue $\lambda$ and its corresponding BS amplitudes
as well as the bound state masses.
Covering the support of $T$-norm is necessary for finding out the
correct value of the fictitious eigenvalue $\lambda$ by our updating
method given in sect.~\ref{MSHBSE}.
The $T'$-norm plays the role for determining whether the state is a
physical state ($T'>0$) or a ghost state ($T'\leq 0$).
The ratio of $T'$- and $T$-norms is essential for checking the error
$\delta M_V$ defined in eq.~(\ref{eq: newton}) to converge the correct
value of the bound state mass.

\begin{figure}[htbp]
\begin{center}
\ \epsfbox{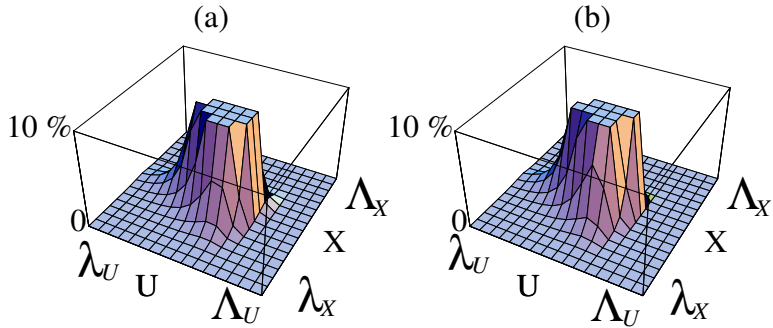}
\vspace{-5pt}
\caption[]{
The supports of a) $T$-norm and b) $T'$-norm for the choice
$m=1.0$ and $M_V=1.2$ with $N_{BS}=17$ for the state $\chi_{1'}$.
The upper $9/10$ of each figures is clipped.
Horizontal axes of $U$ and $X$ are parameterized by
eq.~(\ref{eq: UX}).
Summing up all values on the lattice points give us the norm for a)
and b).
}
\label{fig: support}
\end{center}
\end{figure}

For the evaluation of $M_V^{(1)}$, we need to cover the support of
$\bbra{\chi_{1'}}(-T)\kket{\chi_{1'}}$ if there is no level
annihilation.
When we evaluate $M_V^{(2)}$ we need to cover both the supports of
$\bbra{\chi_{1'}}(-T)\kket{\chi_{1'}}$ and
$\bbra{\chi_{2'}}(-T)\kket{\chi_{2'}}$, and so on.
This is because the excited state $\chi_{n'}$ should be orthogonal
to the lower states $\chi_{1'}$, $\chi_{2'}$, $\cdots$,
$\chi_{(n-1)'}$ in the sense of eq.~(\ref{eq: orthogonality}).%
\footnote{
If we calculate the decay constant, we also need to cover the
support of $T'$-norm in a similar reason.
}
In the following calculations we carefully cover the $T$-norm and
$T'$-norm supports of the first two states, i.e., $\chi_{1'}$ and
$\chi_{2'}$.
We show the typical example of the supports of $T$-norm and $T'$-norm
for the choice $m=1.0$ and $M_V=1.2$ in fig.~\ref{fig: support}.
[The condition $\lon(M_V)=1$ is satisfied by this parameter
choice.]

\subsection{The spectrum of $\lambda$}
\label{sec: spec}

\begin{figure}[bhtp]
\begin{center}
\ \epsfbox{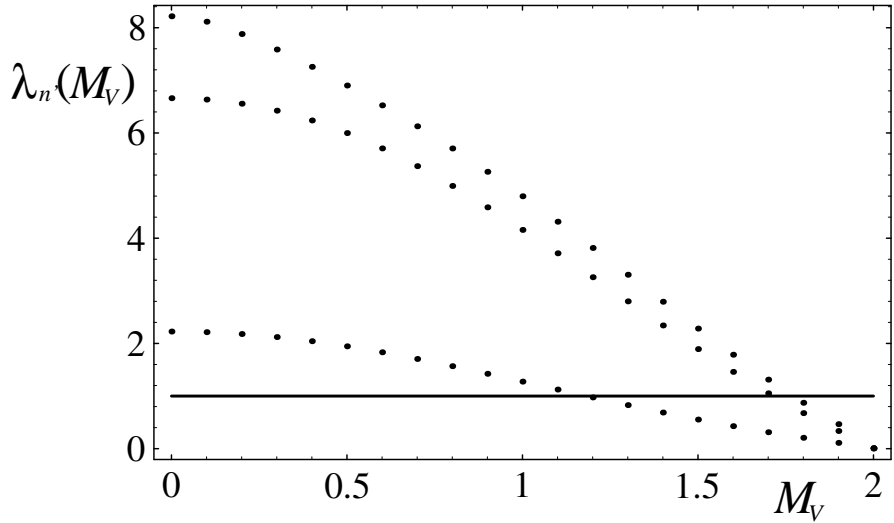}
\vspace{-5pt}
\caption[]{
The functional forms of the smallest three $\lan(M_V)$.
The horizontal line indicates $\lan(M_V)=1$.
We use $N_{BS}=17$.
}
\label{fig: l m=1.0}
\end{center}
\end{figure}

First, we show the functional form of the fictitious eigenvalue
$\lan(M_V)$ with a choice $m=1.0$ in fig.~\ref{fig: l m=1.0}.
This figure obviously shows that the smallest three $\lan(M_V)$'s are
monotonically decreasing functions and there are no level crossing and
level annihilation.
The eigenvalues $\lon(M_V)$ and $\ltw(M_V)$ are the first and the
second to become unity respectively when the
bound state mass increases from zero.
Then, the states $\chi_{1'}$ and $\chi_{2'}$ are corresponding to the
ground state and the first excited state, respectively.
At the free fermion threshold $M_V=2m$ all eigenvalues of $\lambda$
become zero.
There are free-state solutions which satisfy $T\chi=0$ (eq.~(\ref{eq:
eigen HBS}) with $\lambda=0$).
At $M_V=0$ all three slopes of $\lan(M_V)$ become flat smoothly, i.e.,
$d\lan(M_V)/dM_V \rightarrow 0$ as $M_V \rightarrow 0$.
As in eq.~(\ref{eq: diff}), this means that $T'$-norm vanishes
$\bbra{\chi_{n'}}\partial T/\partial M_V \kket{\chi_{n'}}=0$
consistently with the normalization condition
(\ref{eq: amplitude norm}).
Actually, the calculated ratio of $T'$-norm over $T$-norm vanishes at
$M_V=0$.
This result implies that the discrete spectrum at $M_V=0$ is the set
of normal states in the sense of the limit $M_V\rightarrow 0$.

\begin{figure}[bhtp]
\begin{center}
\ \epsfbox{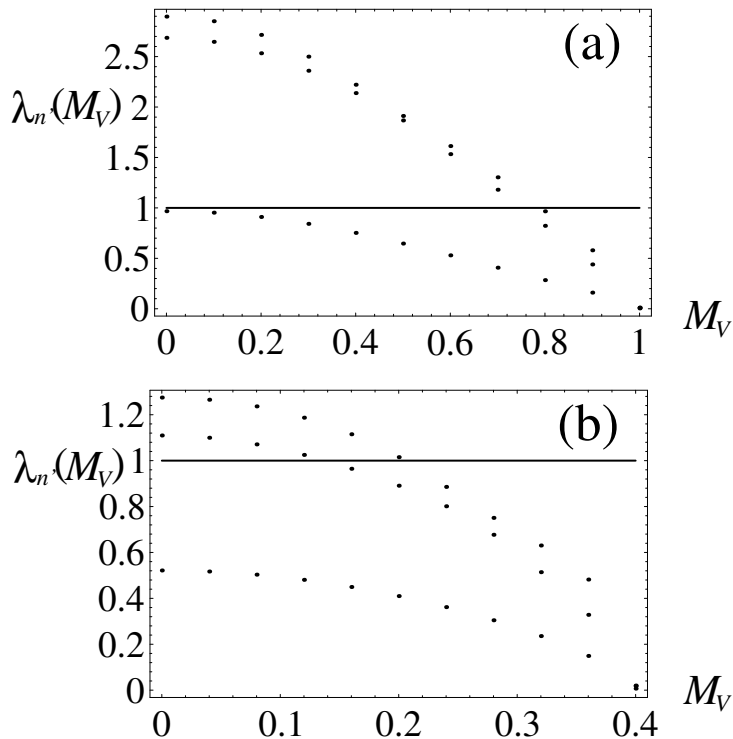}
\vspace{-5pt}
\caption[]{
The plots of the smallest three $\lan(M_V)$.
We use constant fermion masses $m=0.5$ in fig.~(a) and $m=0.2$ in
fig.~(b).
The horizontal lines indicate $\lan(M_V)=1$.
We use $N_{BS}=11$.
}
\label{fig: level anihi}
\end{center}
\end{figure}

Second, we study the cases for smaller fermion masses.
We show the spectrum of the smallest three $\lan(M_V)$ with $m=0.5$
and $0.2$ in fig.~\ref{fig: level anihi}.
When we chose $m=0.5$, the smallest eigenvalue $\lon(M_V)$ takes the
maximum value unity at the vanishing
bound state mass $M_V=0$ as in
fig.~\ref{fig: level anihi} (a).
This means that $m=0.5$ is a critical value for the level
annihilation.
If we chose smaller value of the fermion mass ($m<0.5$), then the
level annihilation of the state $\chi_{1'}$ will occur.
Accordingly, the mass of the ground state changes discretely from zero
to $M_V = 0.73$ around $m\sim 0.5$.
When we chose $m=0.2$ as in fig.~\ref{fig: level anihi} (b), the
smallest eigenvalue $\lon(M_V)$ is always smaller than unity and never
reaches to unity over the range $0\le M_V \le 2m$.
In other words, the equation $\lon(M_V)=1$ has no solution for
$0\le M_V \le 2m$.
This means that the state $\chi_{1'}$ has no correspondence with the
physical solutions of eq.~(\ref{eq: HBS1.b}), i.e., the level
annihilation occurs.
Then, $\chi_{2'}$, $\chi_{3'}$, $\cdots$ are corresponding to
the ground state, first excited state and so on, respectively.
If the constant fermion mass becomes much smaller than $0.2$, the next
smallest eigenvalue $\ltw(M_V)$ will become always smaller than unity
as well as $\lon(M_V)$.
When $m=0$, maximal number of states will fail to have any physical
correspondences.

Here it is important to notice that figs.~\ref{fig: l m=1.0} and
\ref{fig: level anihi} show that the slopes of the smallest three
$\lan(M_V)$ are negative.
Further, we actually observe that all states in consideration have
positive $T$- and $T'$-norms, and this leads, by eq.~(\ref{eq: diff}),
to the fact that the corresponding $\lan(M_V)$ (or $\lambda_n(M_V)$)
are decreasing functions.
This result supports our assumption that all normal solutions have
positive $T$-norm, then all eigenvalues $\lan(M_V)$ are decreasing
functions.

Let us reconsider the level annihilation from the other side.
Since the largest value of $\lan(M_V)$ is realized at the vanishing
bound state mass $M_V=0$, it is enough to calculate the
eigenvalues $\lon$, $\ltw$, $\cdots$ at $M_V=0$ for various constant
fermion masses in order to see whether the level annihilation occurs.
\begin{figure}[bhtp]
\begin{center}
\ \epsfbox{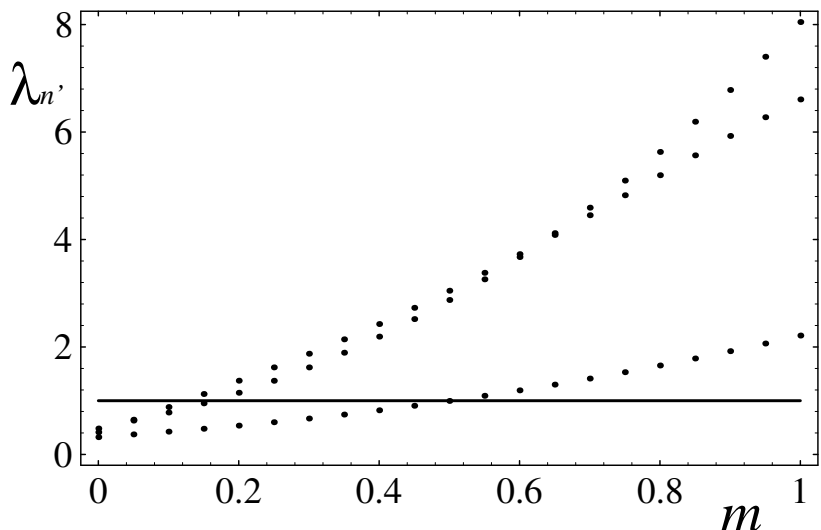}
\vspace{-5pt}
\caption[]{
The smallest three $\lan$ against $m$ at the vanishing
bound state mass $M_V = 0$.
We use $N_{BS}=14$.
At each value of the fermion mass $m$ we distinguish the three points
by the definition of its ordering $\lon \leq \ltw \leq \lth$.
\label{fig: M_V=0}
}
\end{center}
\end{figure}
We show the result in fig.~\ref{fig: M_V=0}.
This figure shows that the first level $\chi_{1'}$ is annihilated at
$m=0.5$, the second level $\chi_{2'}$ are annihilated at $m=0.16$ and
the third level $\chi_{3'}$ at $m=0.13$ and so on.
In other words, the ground state mass is given as a solution of
$\lon(M_V)=1$ for $m>0.5$, while we have to solve $\ltw(M_V)=1$ for
$0.16<m<0.5$.
[The two level crossings with respect to the fermion mass seem to
occur around $m=0.05$ and $0.64$.]
Recall that we are considering only real value of $M_V$.
If we allowed the
bound state mass $M_V$ to take complex value,
then we could always find the solution of $\lan(M_V)=1$, and might
have no level annihilation.
Since we are interested in physical states, this solution is beyond
the scope of this paper.

Next, let us consider the existence of the discrete spectrum, which we
assume so far for simplicity.
It is sufficient to study at the vanishing bound state mass
$M_V=0$.
As is seen from eq.~(\ref{eq: alpha}), the running coupling becomes
constant in the low energy region, which means the low energy behavior
of the model is similar to that of the strong coupling QED.
For a type of BS equation in the fixed coupling case,
there is no discrete spectrum of $\lambda$ (or $\alpha_0\lambda$)
for the tightly bound state, $M_V=0$\cite{Seto}.
So, it is important to investigate whether the discrete spectrum
exists in the running coupling case.
Figure~\ref{fig: M_V=0} shows that the there exists discrete spectrum
for all range of the fermion mass $m>0$.
Clearly the state $\chi_{1'}$ is one of the discrete states:
if we had continuous states, the differences $\ltw-\lon$ and
$\lth-\lon$ would be of order $O(1/N_{BS})$.
But our result shows the order-one differences.
To confirm the existence of the discrete spectrum of the fictitious
eigenvalue $\lambda$,
we use the three choices of lattice sizes $N_{BS}=11$, $14$ and $17$.
We show the values of $\lon$, $\ltw$ and $\lth$ for the fermion mass
$m=1.0$ in table~\ref{tab: discreteness}.
\begin{table}[hbtp]
\begin{center}
\begin{tabular}[t]{|c||c|c|c|}
\hline
$N_{BS}$ & 11   & 14   & 17    \\\hline
$\lon$   & 2.24 & 2.21 & 2.23  \\
$\ltw$   & 6.42 & 6.61 & 6.66  \\
$\lth$   & 7.85 & 8.05 & 8.22  \\
\hline
\end{tabular}
\vspace{0.3cm}
\caption[]{
The dependences of the fictitious eigenvalues $\lon$, $\ltw$ and
$\lth$ on the lattice size for $M_V=0$ and $m=1.0$.
The difference $\lambda_{(n+1)'}-\lan$ is not order
$O(1/N_{BS})$ but order $O(1)$.
}\label{tab: discreteness}
\end{center}
\end{table}
Each $\lan$ changes its value by $O(1/N_{BS})$ as the value of
$N_{BS}$ is changed, while the differences $\lan-\lambda_{m'}$ are
never of such order.
This clearly shows the existence of the discrete spectrum $\{\lan\}$.

Finally, we note other choices for the running coupling.
If we tune $\lambda$ to an appropriate value, say $\lambda_0$,
we obtain the solution of the HBS equation which describes the bound
states in any system with different gauge group and matter contents.
This is realized when we use the improved ladder and constant fermion
mass approximations simultaneously.
When we chose other gauge groups and matter content than $SU(3)$ and
$N_f=3$, the parameter $\alpha_0(N_c\!=\!N_f\!=\!3)\equiv 4\pi/9$
defined in eq.~(\ref{eq: alpha}) is modified.
Since $\alpha_0$ is just an over overall constant in the
right-hand-side of eq.~(\ref{eq: eigen HBS}), we can identify
\be
\lambda = \frac{9}{4\pi}\,\alpha_0 ~,
\ee
and we regard that all figures \ref{fig: l m=1.0},
\ref{fig: level anihi} and \ref{fig: M_V=0} shows the spectrum of
$\alpha_0(M_V)$ scaled by $4\pi/9$.
Figure~\ref{fig: l m=1.0} shows that for the case
$2.23<\lambda_0<6.66$ the equation $\lon(M_V)=\lambda_0$ has no
solution, i.e., the level annihilation occurs and the ground state
mass $M_V^{(1)}$ is given by $\ltw(M_V^{(1)})=\lambda_0$.
For the case $6.66<\lambda_0<8.22$ $M_V^{(1)}$ is given by
$\lth(M_V)=\lambda_0$.
[The condition $\lambda_0>2.23$ corresponds to $N_f>10$ for $N_c=3$.]

\subsection{Bound state mass}
\label{sec: mass}

Using the method explained in section~\ref{MSHBSE},
we evaluate the mass of the vector bound state.
We consider both the fixed and the running coupling cases with
constant fermion mass.
The vector bound state in weak fixed coupling case corresponds to the
orthopositronium, the system of which is well-known both theoretically
and experimentally\cite{positronium}.
We can explicitly check the validity of our method in this system.
Next, we proceed to the running coupling case.

In the weak coupling limit the HBS equation for the orthopositronium
is solved, and we obtain the
famous non-relativistic result:\cite{Tiktopoulos,IZ}
\be
\mvn = 2m - \frac{m\alpha^2}{4n^2} ~,\label{eq: sp posit}
\ee
where $m$ is the pole mass of the fermion and antifermion.
On the other hand, the spectrum of $\lambda$ (or $\lambda\alpha$) for
a small binding energy in eq.~(\ref{eq: eigen HBS}) is given
by\cite{Seto}
\be
\lan(M_V) = \frac{2n}{\alpha}\sqrt{\frac{2m-M_V}{m}}~.
\label{eq: spect}
\ee
Obviously, imposing $\lan(M_V)=1$ in eq.~(\ref{eq: spect}) we obtain
the non-relativistic result (\ref{eq: sp posit}).
The smallest eigenvalue $\lon$ corresponds to the ground state
($n=1$), and the second smallest eigenvalue $\ltw$ corresponds to the
first excited state ($n=2$), and so on.
Furthermore, the binding energy and BS amplitude have no
dependence on the gauge parameter.\cite{FukuiSetoYoshida}
\footnote{
In the fixed strong coupling and constant mass the positronium does
not exist and there are only continuum spectrum and no discrete
spectrum.\cite{Seto}
In the case of the vanishing center-of-mass momentum $q_\mu=0$ the HBS
equation is solved and a strong dependence on the gauge parameter is
observed.\cite{NishiHigashi}
}

Here, we rescale all the dimensionful quantities so as to satisfy the
relation $m\alpha = 1$ fixing the coupling constant as $\alpha=1/137$
when we solve the HBS equation for the positronium.
When the fermion mass becomes small, the support of the $T'$-norm
begins to shift to the infrared region.
This is the reason why we use large value of the fermion mass
compared with the unit scale.

We check whether the numerical calculation reproduce the
relation (\ref{eq: sp posit}) in the weak coupling limit.
The spectrum (\ref{eq: spect}) tells
us that the quantity $\lon/\lan$ should be an integer.
We adjust the bound state mass $M_V$ as $\lon(M_V)=1$:
we update $M_V$ until the difference $\delta M_V$ given in
eq.~(\ref{eq: newton}) satisfies $|\delta M_V|<m\alpha^4$.
We show the spectrum of the eigenvalues $\lan$ with $N_{BS}=11$, $14$
and $17$ in table~\ref{tab: spectrum of lambda}.
\begin{table}[hbtp]
\begin{center}
\begin{tabular}[t]{|c||c|c|c|}
\hline
$N_{BS}$
        &   11 &      14 & 17      \\\hline
$\lon/\ltw$
        & 2.12 &    2.10 & 2.09  \\
$\lon/\lth$
        & 3.13 &    3.16 & 3.17  \\
$\lon/\lambda_{4'}$
        & 3.37 &    3.32 & 3.29  \\
$\lon/\lambda_{5'}$
        & 4.36 &    4.39 & 4.38  \\
\hline
\end{tabular}
\vspace{0.3cm}
\caption[]{
The spectrum of eigenvalues $\lan$ for $N_{BS}=11$, $14$
and $17$ when the bound state mass is adjusted such that the
largest eigenvalue $\lambda_{1'}$ is equal to unity.
This should be compared with the result in the
non-relativistic limit, $\lon/\lan=n$.
}\label{tab: spectrum of lambda}
\end{center}
\end{table}
The results show that $\lon/\lan$ appears as an integer for the first
several $n$'s within about 10\% errors.
As for the quantum number $n'$ larger than five, $\lon/\lan$ suffers
large numerical errors.
This simply comes from the fact that we do not cover the supports of
such higher radial excited states.
We calculate the binding energies of the first three states.
The binding energy is defined by $B^{(n)} = 2m - \mvn$ and is given in
table~\ref{tab: binding energy} with $N_{BS} = 11$, $14$ and $17$.
\begin{table}[hbtp]
\begin{center}
\begin{tabular}[t]{|c||c|c|c|}
\hline
$N_{BS}$
        & 11     & 14     & 17    \\\hline
$B^{(1)}$ [Ryd]
        & 0.513  & 0.489  & 0.479  \\
$B^{(2)}$ [Ryd]
        & 0.121  & 0.116  & 0.115  \\
$B^{(3)}$ [Ryd]
        & 0.0544 & 0.0515 & 0.0505 \\
\hline
\end{tabular}
\vspace{0.3cm}
\caption[]{
The biding energies for the ground state ($B^{(1)}$),
the first excited state ($B^{(2)}$) and
the second excited state ($B^{(3)}$)
with $N_{BS} = 11$, $14$ and$17$.
The values are in the units of the Rydberg energy,
$m\alpha^2/2$.
}\label{tab: binding energy}
\end{center}
\end{table}
This result should be compared with that in the non-relativistic
limit: $B^{(n)} \mbox{[Ryd]}
=1/(2n^2)$ ($= 0.5$, $0.125$, $0.0556$, $\ldots$), and shows that our
method works well.

\begin{figure}[bhtp]
\begin{center}
\ \epsfbox{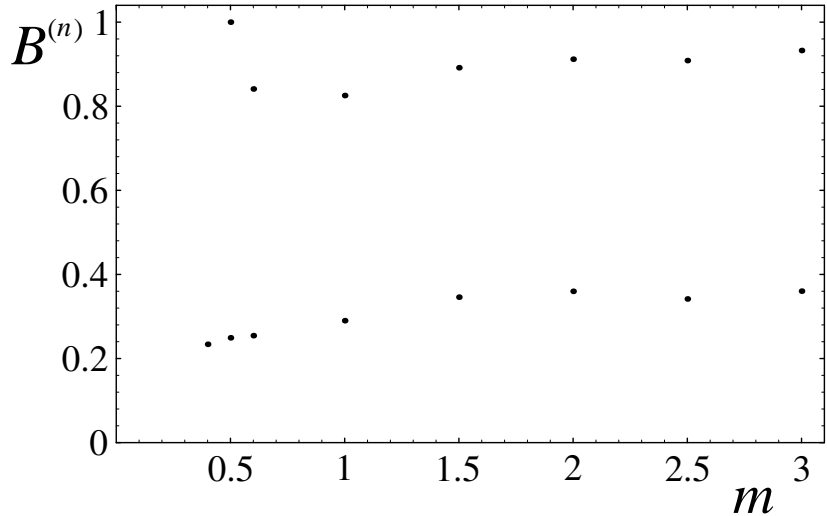}
\vspace{-5pt}
\caption[]{
The binding energies $B^{(1)}$ and $B^{(2)}$ for various constant
fermion masses.
The upper half data are of $B^{(1)}$ and the lower half are of
$B^{(2)}$.
}
\label{fig: M_V vs m}
\end{center}
\end{figure}

Now that, we proceed to the running coupling case.
We update the bound state mass $M_V$ until the difference
$\delta M_V$ given in eq.~(\ref{eq: newton}) satisfies $|\delta
M_V|<10^{-2}$.
As studied in section~\ref{sec: spec}, the ground state mass is given
by the solution of $\lon(M_V)=1$ for $0.5<m$ and by $\ltw(M_V)=1$ for
$0.16<m<0.5$.
In the latter case, the ground and first excited states are given by
$\chi_{2'}$ and $\chi_{3'}$, respectively, due to the level
annihilation.
We show the binding energies ($B^{(n)} \equiv 2m - \mvn$) of the
ground and first excited states for various fermion masses in
fig.~\ref{fig: M_V vs m}.
This figure shows that $B^{(1)}=2m$ at $m=0.5$, i.e., $\chi_{1'}$ is
massless state.
At $m=0.4$ the state $\chi_{1'}$ and its eigenvalue $\lon$ have no
physical correspondence due to the level annihilation, and
the binding energy of only the ground state is calculated from
$\ltw(M_V)=1$.
As a result, the binding energy of the ground state changes
discretely from $1.0$ ($=2m$) to $0.25$ around $m\sim 0.5$.
We make some comments here.
The binding energies $B^{(1)}$ and $B^{(2)}$ approach to the values
$B^{(1)} \rightarrow 1$ and $B^{(2)} \rightarrow 0.4$ asymptotically
for large fermion mass.
It is plausible that the binding energies of a heavy bound states do
not depend on the mass of their heavy constituents thanks to the
decoupling theorem\cite{decouple}.
We may also read off the values of fermion masses
with which we obtain massless bound states
from fig.~\ref{fig: M_V=0}.

\section{Discussions}
\label{Summary and Discussion}
\reseteqnum

As discussed in the Introduction, ideally we are better to solve the
SD equation in the same approximation as the BS equation in order to
obtain the full propagator (or mass function).
For the purpose that we study the numerical method for solving the HBS
equation for the massive bound state, we use the constant, i.e., tree
level, mass $\Sigma(x) = m$ in this paper.
It is true that this approximation is good in the weak coupling QED
and retains the non-relativistic result.
For applying our method to the real QCD case, however, we should use
the mass function in the fermion propagator.
Even when we consider the heavy quarkonia, the quantum corrections to
the mass function have the same order as the binding energy.
Especially in the case of the light quarkonia ($\rho$ meson) the
effect of the mass function becomes more important.
As a result, the running effect of mass function is always
important for any quarkonia, and is same order of magnitude as the
binding force.

In the real QCD there exists the states with negative binding energy
(for highly excited states) because of the confining potential.
Our choice of the running coupling does not generate the confining
force, while a suitable form of the running coupling generates
it.
The non-relativistic quark model with the Richardson potential gives a
good result for the low-lying spectrum of the heavy
quarkonia.\cite{Richardson}
In the constant mass approximation the $q\overline q$ free quark
threshold opens at $q^2=(2m)^2$.%
\footnote{
If we take into account of the running effect of the mass function by
solving the SD equation, the resultant quark propagator has no pole
for the real momentum.\cite{FukudaKugo,AtkinsonBlatt}
}
Thus the states with negative binding energy are unstable.
We observe that infinitely many number (the order of total lattice
points) of states are lying in the region $M_V<2m$.
We consider only the states which have positive binding energy.
Only a first few state will be relevant in the constant mass
approximation.

In this paper, we study the the solutions of the homogeneous
Bethe-Salpeter equation in the constant mass and improved ladder
approximations.
We leave the application of our method to the quarkonia for future
work.

\vspace{1cm}
\appendix
\begin{center}
\Large\bf Appendix
\end{center}

\section{Schwinger-Dyson Equation}
\label{SD}
\reseteqnum

In this appendix, we briefly review how to derive the fermion mass
function in the complex plane.
Here, following ref.~\cite{KY},
we perform the analytic continuation to the SD equation in the
integral form.

In order to solve the Bethe-Salpeter (BS) equation we need
a fermion propagator $S_F(p)$, which is determined by the
Schwinger-Dyson (SD) equation.
When the ladder approximation is adopted for the BS equation,
it is natural to use the same approximation for the SD equation.
In the strong interaction the mass function of the quark propagator
receives large quantum correction of the same order $\lqcd$ as that of
binding force.

The SD equation in the (improved) ladder approximation is given by
\be
i\SI(p) = \slp - m - \intdk~ C_2g^2(p,k)
\frac{1}{-l^2}\left(g_{\mu\nu}-\frac{l_\mu l_\nu}{l^2}\right)~
\gmu~iS_F(k)~\gnu ~, \label{eq: SD1}
\ee
where $l_\mu = (p-k)_\mu$, $C_2=(N_c^2-1)/(2N_c)$ is a second Casimir
invariant and $m$ is the bare mass of the fermion.
The Landau gauge is adopted for the propagator of the gauge boson.
When we consider QCD, we use the running coupling in the SD equation
(\ref{eq: SD1}).
Here for definiteness we adopt the Higashijima-Miransky
approximation\cite{Higashijima,Miransky} to the running coupling,
i.e., $g^2(p,k) = g^2(\max(-p^2,-k^2))$.
In considering QED we use fixed coupling $g^2(p,k) = e^2$ and define
$C_2 = 1$.

In general the fermion propagator is expanded by two scalar functions
$A$ and $B$:
\be
i\SI(p) = A(-p^2)\slp - B(-p^2) ~.\label{eq: fermion prop}
\ee
The mass function $\Sigma(z)$ is defined by $\Sigma(z) = A(z)/B(z)$.
Substituting eq.~(\ref{eq: fermion prop}) into the SD equation
(\ref{eq: SD1}) we find $A(z) = 1$ and
\be
  \Sigma(z) = m +
  \frac{3C_2}{16\pi^2}
  \int_0^\infty\!\!ydy\,
  \frac{g^2(\max(z,y))}{\max(z,y)}\,
  \frac{\Sigma(y)}{y+\Sigma^2(y)} ~.
\label{eq: SD2}
\ee

Now, let us discuss the problem i) stated in
sec.~\ref{Introduction}.
When we solve the homogeneous BS equation for the massive bound state,
we have to use the mass function on the complex plane.
The mass function $\Sigma(z)$ is needed along the set of parabolic
curves $z=(u\mp iM_V/2)^2 + x^2$, where $M_V$ is the mass of the bound
state.
We can restrict ourselves to the case ${\rm Im}~z >0$ because of
$\Sigma(z)^*=\Sigma(z^*)$.
It is not difficult to calculate such a mass function.
In refs.~\cite{FukudaKugo,AtkinsonBlatt}
the SD equation
(\ref{eq: SD2}) is converted into the differential equation for
carrying out the analytic continuation of the mass function.
On the other hand, we can easily solve the SD equation (\ref{eq: SD2})
as the integral equation itself by the iteration.\cite{KY}
After carrying out the analytic continuation, the SD equation becomes
\bea
\lefteqn{
  \Sigma(z) = m + \frac{3C_2}{16\pi^2}
  \Bigg[\,
    \int_0^{-M_V^2/4}\!\!dy\,\frac{yg^2(z)}{z}
}\nn
&& \qquad +
    \mathop{\int}_{C(-M_V^2/4,z)}\!\!dy\,
    \frac{yg^2(z)}{z}
    + \mathop{\int}_{C(z,\infty)}\!\!dy\, g^2(y)
  \,\Bigg]\,
  \frac{\Sigma(y)}{y+\Sigma^2(y)} ~,
\label{eq: SD3}
\eea
where $C(a,b)$ is the contour from the point $a$ to $b$ and is lying
on the parabolic curve with ${\rm Im}~z>0$.
To say more completely, we should take account of the singularity
of the running coupling.
We can solve the SD equation (\ref{eq: SD3}) by the iteration on
the parabolic curve as usual.
Then, we obtain the mass function on the complex plane from
eq.~(\ref{eq: SD3}).

\clearpage
\begin{center}
\Large
Acknowledgments
\end{center}

We would like to thank Taichiro Kugo and Joseph Schechter for
helpful discussions and comments.
We also thank Ken Nakanishi for a comment.

%
%
\newcommand{\PR}[1]{{Phys.~Rev.}~{\bf #1}}
\newcommand{\PRD}[1]{{Phys.~Rev.~D}~{\bf #1}}
\newcommand{\PRL}[1]{{Phys.~Rev.~Lett.}~{\bf #1}}
\newcommand{\PRep}[1]{{Phys.~Rep.}~{\bf #1}}
\newcommand{\PL}[1]{{Phys.~Lett.}~{\bf #1}}
\newcommand{\PLB}[1]{{Phys.~Lett.~B}~{\bf #1}}
\newcommand{\MPL}[1]{{Mod.~Phys.~Lett.}~{\bf #1}}
\newcommand{\NP}[1]{{Nucl.~Phys.}~{\bf #1}}
\newcommand{\SJNP}[1]{{Sov.~J.~Nucl.~Phys.}~{\bf #1}}
\newcommand{\AP}[1]{{Ann.~Phys.}~{\bf #1}}
\newcommand{\PTP}[1]{{Prog.~Theor.~Phys.}~{\bf #1}}
\newcommand{\NC}[1]{{Nuovo~Cim.}~{\bf #1}}
\newcommand{\JMP}[1]{{J.~Math.~Phys.}~{\bf #1}}
\newcommand{\ibid}[1]{{ibid. \bf #1}}

\end{document}